\documentclass{ptephy_v1}

\usepackage{amsmath}
\usepackage{geometry}
\usepackage{graphics}
\usepackage{multirow}
\usepackage{url}
\usepackage{float}

\usepackage{color}

\begin{document}

\title{Angular correlation of the two gamma rays produced in the thermal neutron capture on gadolinium-155 and gadolinium-157}

\author[1,2]{Pierre Goux}
\author[1,2]{Franz Glessgen}
\author[1,3,8]{Enrico Gazzola}
\author[1,9]{Mandeep Singh Reen}
\author[2]{William Focillon}
\author[2,5,*]{Michel Gonin}
\author[1]{Tomoyuki Tanaka}
\author[1]{Kaito Hagiwara}
\author[1,10]{Ajmi Ali}
\author[1,6]{Takashi Sudo}
\author[1]{Yusuke Koshio}
\author[1,*]{Makoto Sakuda}
\author[3]{Gianmaria Collazuol}
\author[4]{Atsushi Kimura}
\author[4]{Shoji Nakamura}
\author[4]{Nobuyuki Iwamoto}
\author[4,*]{Hideo Harada}
\author[7]{Michael Wurm}
\affil[1]{Department of Physics, Okayama University, Okayama 700-8530, Japan }
\affil[2]{D\'epartement de Physique, \'Ecole Polytechnique, IN2P3/CNRS, 91128 Palaiseau Cedex, France}
\affil[3]{University of Padova and INFN, Italy}
\affil[4]{Japan Atomic Energy Agency, 2-4 Shirakata, Tokai, Naka, Ibaraki 319-1195, Japan}
\affil[5]{ILANCE, CNRS – University of Tokyo International Research Laboratory, Kashiwa, Chiba 277-8582, Japan}
\affil[6]{Research Center for Nuclear Physics (RCNP), Osaka University, 567-0047 Osaka, Japan}
\affil[7]{Institut f\"ur Physik, Johannes Gutenberg-Universit\"at Mainz, 55128 Mainz, Germany}
\affil[8]{Present address: Finapp srl, 35036 Montegrotto Terme, Padua, Italy}
\affil[9]{Present address: Department of Physics, Akal University, Punjab 151302, India}
\affil[10]{Present address: Department of Physics, University of Winnipeg, Manitoba, Canada}
\affil[ ]{\email{sakuda-m@okayama-u.ac.jp, michel.gonin@polytechnique.edu, harada.hideo@jaea.go.jp}}

\begin{abstract}
The ANNRI-Gd collaboration studied  in detail the single  $\gamma$-ray spectrum produced from the thermal neutron capture on $^{155}$Gd and $^{157}$Gd in our previous publications. Gadolinium targets were exposed to a neutron beam provided by the Japan Spallation Neutron Source (JSNS) in J-PARC, Japan. 
In the present analysis, one new additional coaxial germanium crystal was used in the analysis in combination with the fourteen  germanium crystals in the cluster detectors to study the angular correlation of the two $\gamma$ rays emitted in the same neutron capture. We present for the first time angular correlation functions for two $\gamma$ rays produced during the electromagnetic cascade transitions in the (n, $\gamma$) reactions on $^{\rm 155}$Gd and $^{\rm 157}$Gd.  As expected, we observe the mild angular correlations for the strong, but rare  transitions from the resonance state to the two energy levels of known spin-parities.  Contrariwise,  we observe negligibly small angular correlations for arbitrary pairs of two  $\gamma$ rays produced in the majority of cascade transitions from the resonance state to the dense continuum states. 
   
\end{abstract}

\subjectindex{C30, C43, D03, F20, H20}

\maketitle

\section{Introduction}

 The gadolinium (Gd) nucleus is one of the few stable nuclei (Cd, Sm, Gd) featuring unusually large cross sections and resonance enhancements for the thermal neutron capture~\cite{Mughabghab2006, Leinweber, Choi, nTOF}. The two gadolinium isotopes $^{157}$Gd and $^{155}$Gd possess the largest neutron capture cross sections among the stable nuclei~\cite{Mughabghab2006}.
 The element has been used as a neutron absorber in liquid-scintillator-based detectors for neutrino oscillation experiments~\cite{Dchooz, RENO, DayaBay, NEOS, STEREO, Neutrino4, DANSS, SterileCombine, JSNS2},  a neutrino-flux monitor experiment~\cite{PANDA}, and even medical science~\cite{GdNCT}. 
 The application of  Gd-loaded detectors for tagging neutrons has been recently extended to direct dark matter search experiments~\cite{LZ, Xenon}.  The identification of neutrons produced from the inverse beta decay with large efficiency is crucial for the detection of  Supernova Relic Neutrinos (SRN) in a Gd-loaded water Cherenkov detector like Super-Kamiokande~\cite{Vagins, EGADS, SkGd}. Upon neutron capture, the  Gd isotopes $^{157}$Gd and $^{155}$Gd release cascade of $\gamma$ rays with a total energy of 7937 keV for $^{158}$Gd and 8536 keV for $^{156}$Gd.  Due to the Cherenkov threshold applying for the detection of the multiple Compton electrons produced by these $\gamma$ rays in the Super-Kamiokande detector, a precise knowledge and understanding of the cascade $\gamma$-ray energies is absolutely necessary in order to model  the neutron capture efficiency with Monte Carlo simulations. 

 In our previous publications~\cite{Hagiwara, Tanaka}, we reported  measurements of the single $\gamma$-ray spectra  produced from the thermal neutron capture on targets comprising a natural Gd film and gadolinium oxide powders enriched with  $^{\rm 155}$Gd and $^{\rm 157}$Gd, where we used the two cluster detectors of the ANNRI spectrometer at J-PARC, covering 15\% of the solid angle with respect to the target. Moreover, we showed that our Monte Carlo simulation (ANNRI-Gd Model) agreed with our measured spectra reasonably well. 
 We first identified the prominent photopeaks above 5 MeV in the single spectrum and found the secondary transitions associated with each primary photopeak. We listed those  12 and 15 primary $\gamma$ rays for $^{\rm 155}$Gd and $^{\rm 157}$Gd, respectively,  and also identified the secondary $\gamma$ rays.  Those 'discrete' $\gamma$ rays constitute 3-7$\%$ of the total  $\gamma$ rays. However, most of the $\gamma$ rays result from the dense 'continuum' states. 
 
 In the present paper, we report on  the angular correlations between the two $\gamma$ rays  for some selected discrete and continuum transitions in the $^{\rm 155}$Gd and $^{\rm 157}$Gd(n, $\gamma$) reactions. One new additional coaxial germanium crystal was introduced in the analysis in combination with the fourteen  germanium crystals in the cluster detectors to study the angular correlation of the two $\gamma$ rays emitted in the same neutron capture. Although  the solid angle covered by the single coaxial detector for a $\gamma$ ray  emitted in the target is  only 1.0\% and that covered by the cluster detectors is 15\%, the coaxial detector has played an essential role in the analysis and it has made it possible to present the result of the angular correlation of the two $\gamma$ rays over the entire region of  $cos\theta_{12}$:  while the range of  $cos\theta_{12}$ for the angle $\theta_{12}$ between the two $\gamma$ rays measured by the cluster detectors is limited to  -1$<cos\theta_{12}<$-0.6 and 0.6$<cos\theta_{12}<$1.0,  a new range -0.4$<cos\theta_{12}<$0.4 has been covered by measuring the angle $\theta_{12}$ between the one coaxial detector and another crystal of the cluster detectors. 
 We show a comparison between our data and the expected angular correlations. The theoretical calculations for the electromagnetic cascade transitions and the angular correlation function can be found elsewhere~\cite{Frauenfelder, Biedenharn, Rose}. 
  
It is not only essential for many detectors using gadolinium~\cite{Dchooz, RENO, DayaBay, NEOS, STEREO, Neutrino4, DANSS, SterileCombine, JSNS2,PANDA,LZ, Xenon,SkGd} to improve their detector simulations of the energy spectra of the $\gamma$ rays for the high accuracy analysis, but also very important to understand the basic feature of the angular correlation of the two $\gamma$ rays produced from the discrete and continuum transitions in the thermal neutron capture reactions. 
 Recently, a Monte Carlo simulation called the FIFRELIN code~\cite{FIFRELIN} has been developed for the STEREO experiment~\cite{STEREO} which takes into account the angular correlations in cascade transitions. Thus, our new data of the angular correlations measured in the $^{\rm 155}$Gd and $^{\rm 157}$Gd(n, $\gamma$) reactions are expected to improve and validate the detector simulation.
\newline

\begin{figure}[!htbp]
  \begin{center}
	\includegraphics[width=0.8\linewidth]{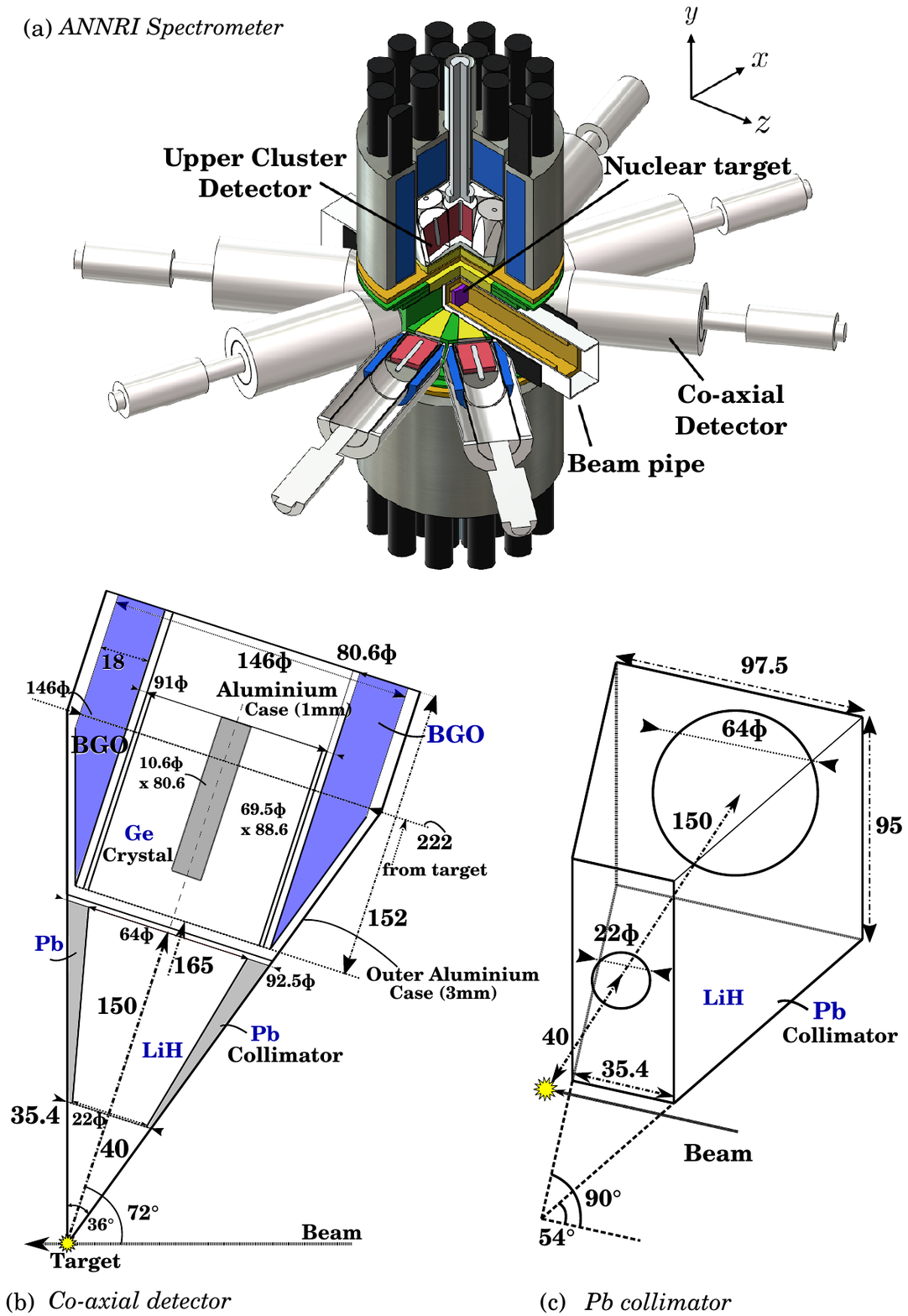}
  \caption{\label{fig::mafigure}ANNRI Spectrometer.}
	\end{center}
\end{figure}

\section{Experiment}
\subsection{Experiment And Data Collection}

The data presented in this analysis were recorded  in December, 2014, with the ANNRI germanium (Ge) spectrometer at the the Materials and
Life Science Experimental Facility (MLF) of J-PARC. The MLF provides a pulsed neutron beam with energies from a few meV up to 100 keV.  Gadolinium oxide powder targets (Gd$_2$O$_3$)  enriched with ${}^{155}$Gd (91.85\%) and ${}^{157}$Gd (88.4\%) were placed inside the ANNRI Ge spectrometer,  which consists of two basic parts: two cluster detectors placed perpendicular to the beam pipe, and eight coaxial detectors placed in a horizontal plane containing the beam pipe and the gadolinium target~\cite{Kimura2012, Kin2011, Kino2011, Kino2014}. The two cluster detectors and  only one of the coaxial detectors were operational during the experiment. A complete description of our experiment and the analysis method, using the two cluster detectors, which consisted of a total of 14 Ge crystals,  can be found in our previous publications~\cite{Hagiwara, Tanaka}.  

In the present analysis, we analysed the data recorded by the single coaxial detector in the horizontal plane in addition to the two cluster detectors. The overall view of the ANNRI spectrometer including both the cluster detectors and the coaxial detectors is shown in Fig.1(a); the detailed geometry of the single coaxial detector and its lead shield is shown in Fig.1(b) and Fig.1(c). The energy threshold for $\gamma$ detection of the coaxial detector was about 300 keV. The other seven coaxial crystal detectors were in repair and were not operational during our experiment. 
The single coaxial detector covers  1.0\%  and the cluster detectors cover about 15\% of the 4$\pi$ solid angle for a $\gamma$ ray from the target. As already stressed in the introduction, the present measurement of the angular correlation of the two $\gamma$ rays over the entire region of  $cos\theta_{12}$ has been made possible by the combination of 
the coaxial detector and the cluster detectors. 


\subsection{Calibration}
The coaxial detector is self-contained with an individual aluminum protective case.  In addition to the protective layer, it is also protected by a lead collimator to reduce the solid angle of $\gamma$ rays produced outside the gadolinium target. LiH was filled inside the lead collimator to reduce the neutron background. With the use of cluster detectors alone, the angular correlation measurements would have been quite limited. The addition of the one  coaxial detector allows for a much greater angular coverage, and substantially larger statistics for the angular correlation analysis. The efficiency of the coaxial detector was estimated with the same method as used for the cluster detectors. This method is described in detail in our previous publications~\cite{Hagiwara, Tanaka}.  In brief, we used radioactive calibration sources, e.g. $^{60}$Co, $^{137}$Cs and $^{152}$Eu, as well as the prompt $\gamma$ rays produced by neutron capture reaction $^{35}$Cl($n, \gamma)^{36}$Cl in the energy range between 0.1 MeV and 9 MeV. To calculate the $\gamma$-ray detection efficiency of the coaxial detector, we  divide the number of $\gamma$ rays detected within the photopeaks by the number of $\gamma$ rays expected due to the solid angle of our detector, corrected by the lifetime of our data acquisition.  \newline

The efficiency values obtained for the coaxial detector are shown in Fig.2  as a function of  $\gamma$-ray energy and are compared to our Geant4 Monte Carlo simulation (dashed-dotted curve) that includes the full geometry and materials of the ANNRI detector. The absolute normalization of our data to the simulation was obtained using  the 7414 keV line of the capture reaction $^{35}$Cl($n, \gamma)^{36}$Cl.  The size of the error bars is determined by the statistics of the data and the Monte Carlo simulation. The agreement between our calibration data and the detector simulations (dashed-dot curve) is satisfactory.  

\begin{figure}[ht!]
  \begin{center}
	\includegraphics[width=10.0cm,scale=1.0]{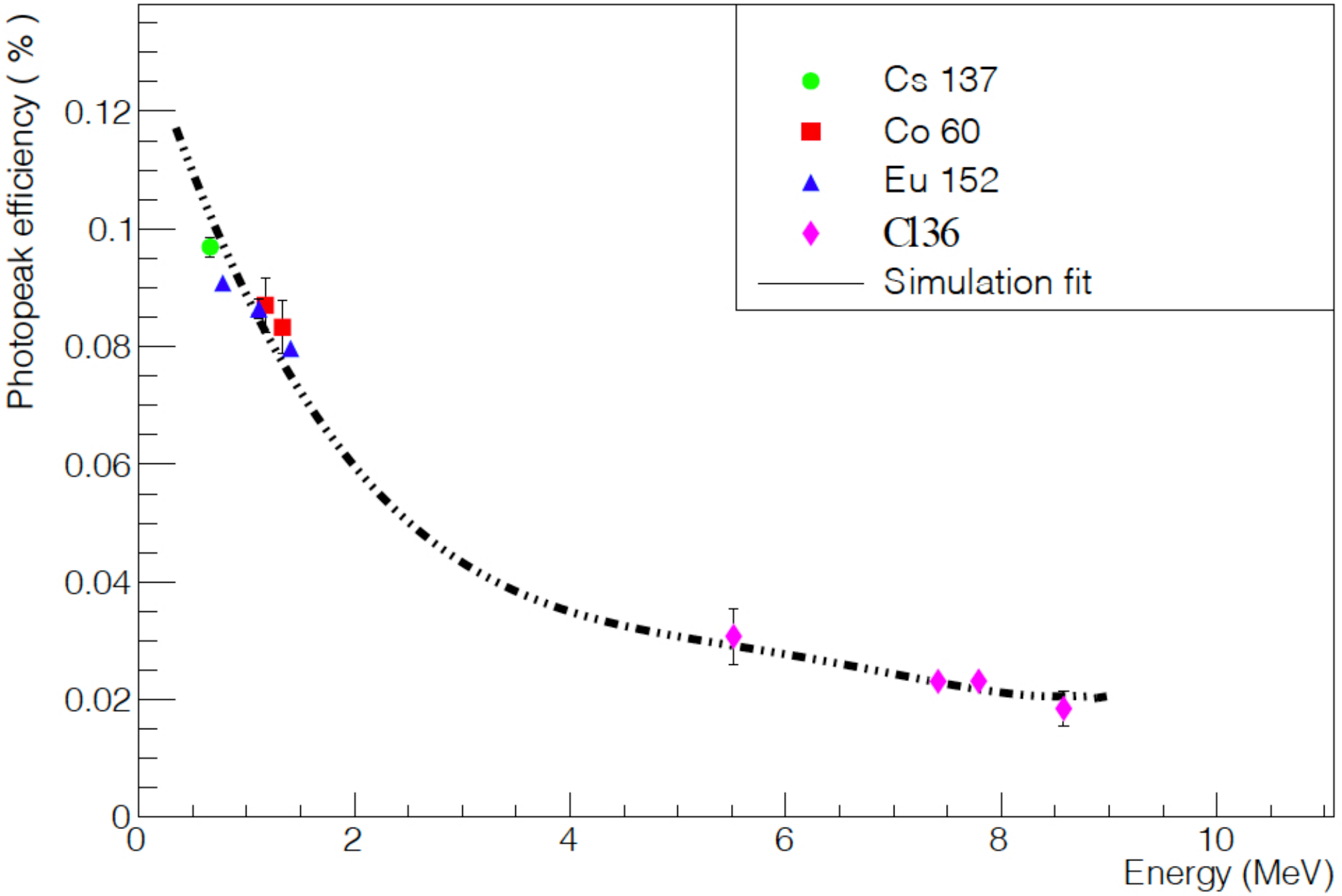}
  \caption{\label{fig1} Efficiency measurements for the coaxial detector compared with simulations(dashed-dot curve).}
	\end{center}
\end{figure}

In the previous publications~\cite{Hagiwara, Tanaka}, we studied the uniformity of the counting rate measured by each crystal of the cluster detectors using the data of radioactive calibration sources. Here, we present a new analysis of the uniformity of both the two clusters and the coaxial detector using the prominent photopeaks produced by exposing the gadolinium targets. For a given photopeak, we calculated for each crystal the ratio of the number of raw data events divided by the expected numbers after taking into account the efficiency, the solid angle and the relative intensity of each photopeak. Fig.3 shows the results for $^{158}$Gd (top) and $^{156}$Gd (bottom). In the histograms, the detector number 0 represents the coaxial detector while the numbers 1 to 14 correspond to the crystals of the two cluster detectors.  The figures  show a very good uniformity between the 15 detectors over an energy range of 1 MeV up to 7 MeV. The variation of the ratios by about 10$\%$ is taken as a measure of  the systematic uncertainties of the counting efficiencies. This uniformity of the measured rates over the two cluster detectors and the coaxial detector is an essential prerequisite for the present analysis of the angular correlations.


\begin{figure}[!htbp]
  \begin{center}
	\includegraphics[width=12.0cm,scale=1.0]{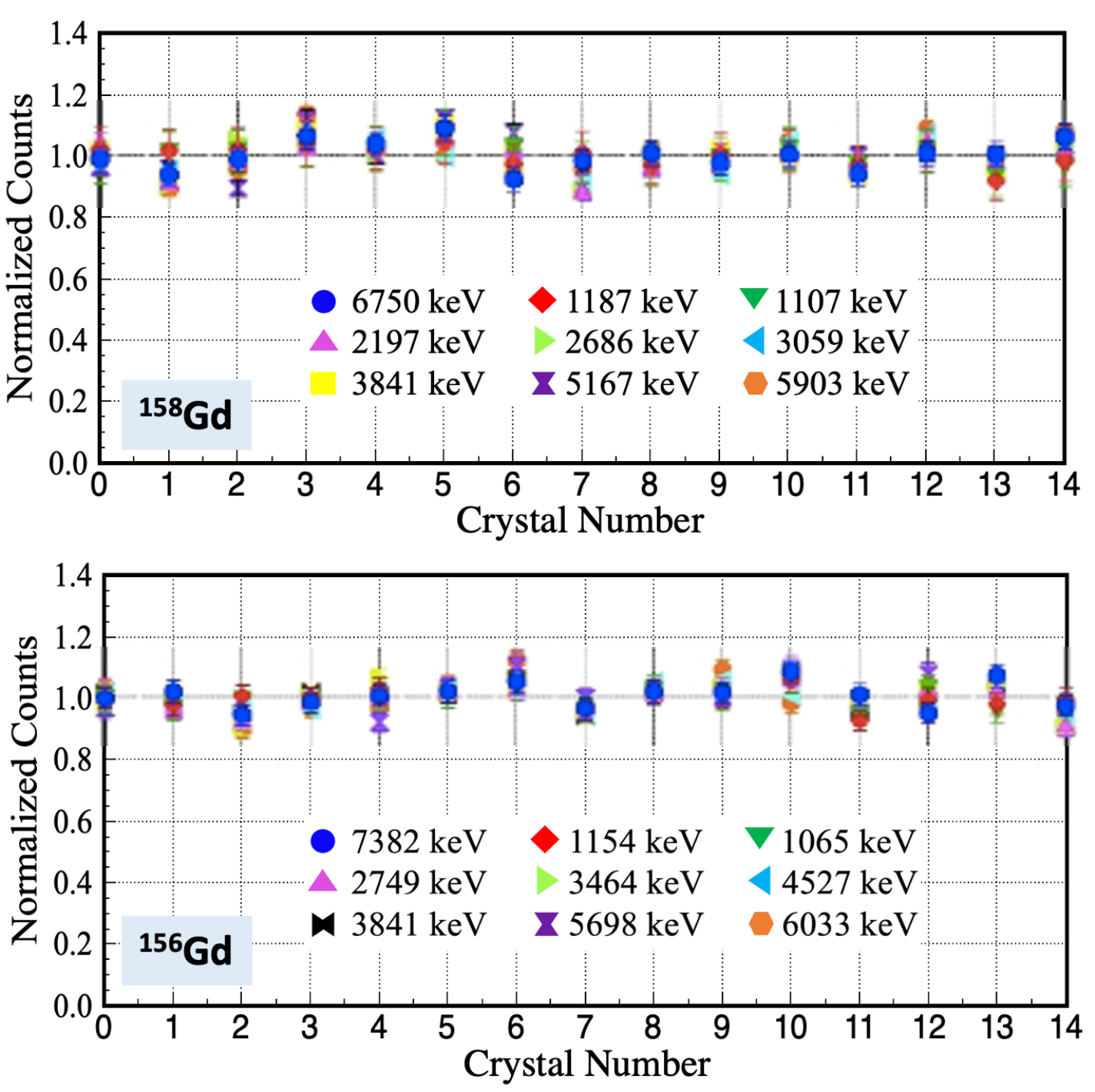}
  \caption{\label{fig2}Normalized counts of various photopeaks for different energy range and for each crystal of the 15  detectors. The top and bottom plots are for the counting rate of various $\gamma$ rays from $^{158}$Gd and $^{156}$Gd, respectively.}
	\end{center}
\end{figure}

\subsection{Data selection and the definition of the angular correlation function  $W(\theta)$}

  As in the previous publications~\cite{Hagiwara, Tanaka},  we classify events by assigning a multiplicity value M and a hit value H to each event.  We defined the multiplicity M  as the combined number of isolated sub-clusters of hit Ge crystals at the upper and the lower  cluster detectors. If the coaxial crystal is hit,  the values of M and  H are both increased by 1, since the hit  is always isolated. The multiplicity M represents the number of observed $\gamma$ rays and the hit value H represents  the total number of Ge crystals hit in the event. We select events categorized as M2H2 and M2H3  to  study the angular correlations of two $\gamma$ rays. \\

\begin{figure}[!htbp]
  \begin{center}
	\includegraphics[width=12.0cm,scale=1.0]{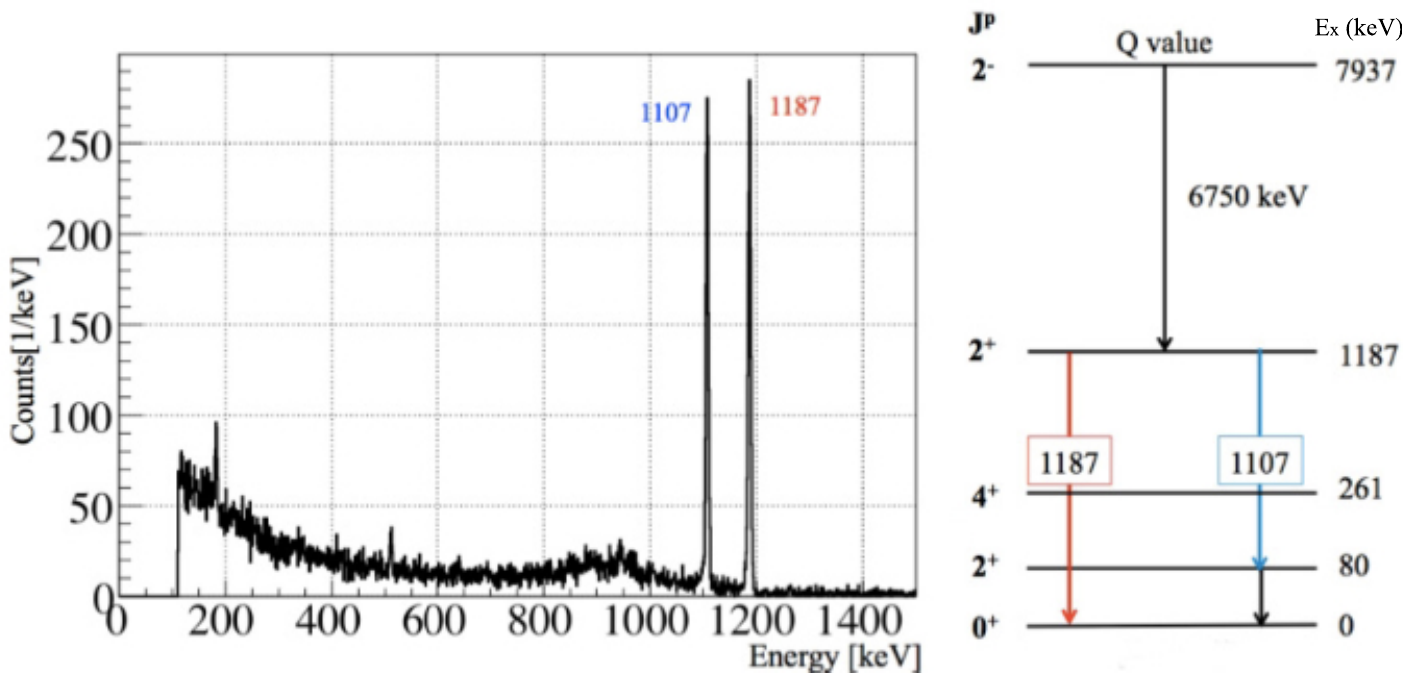}
  \caption{\label{fig3} Energy spectra of $\gamma$ ray measured by the ANNRI detector (left) and illustration of the two- and three-step $\gamma$-ray cascades (right) of $^{158}$Gd after thermal neutron capture.}
	\end{center}
\end{figure}

We define the angular correlation function  $W(\theta_{12})$ using a sample of 
two $\gamma$ rays detected by the two crystals $(i,j)$,  where $\theta_{12}$ is the angle between the two hit crystals $(i,j)$~\cite{Frauenfelder, Biedenharn, Rose}. 
For thermal neutron capture on $^{158}$Gd,  a typical process producing two $\gamma$ rays  is a two-step or three-step cascade transition in the deexcitation of the initial state of 7937 keV ($2^-$). As illustrated in Fig.\ref{fig3} (right), two $\gamma$ rays of 6750 keV and 1187 keV, or three $\gamma$ rays of 6750 keV, 1107 keV and 80 keV, are produced in these transitions. Fig.\ref{fig3}(left) shows the energy of the second  $\gamma$ ray in case the first $\gamma$ ray of 6750 keV is tagged, using the M2H2 (multiplicity 2) sample.  Two peaks corresponding to the $\gamma$ rays of 1187 keV and 1107 keV are clearly seen, while 80 keV below the energy threshold is not measured.  

The observed number $N_{ij}(\theta_{12})$ of two $\gamma$-ray events with energies   $E_1$  and $E_2$  detected in crystals $i$  and $j$  can be denoted  as  
\begin{eqnarray}
N_{ij}(\theta_{12})= N_0 r_{L,ij} \epsilon_{i}(E_1)  \epsilon_{j}(E_2)  W(\theta_{12}),
\end{eqnarray}
where $N_0$ is the number of two $\gamma$-ray events produced at the target, 
$r_{L,ij}$ is the dead time correction factor for the crystal pair $(i,j)$, which typically is on the order of 90\%,  and  $\epsilon_{i}(E_1)$ and $\epsilon_{j}(E_2)$) are the single photopeak efficiency of  the crystals $i$  and $j$ for $\gamma$-ray  energies $E_1$  and $E_2$, respectively, and  $W(\theta_{12}) $ is the angular correlation function between the two $\gamma$ rays.  
If there is no angular correlation, then the angular correlation function is uniform, $W(\theta_{12})=1.0 $, with respect to $cos \theta_{12}$. The angular correlation function $W(\theta_{12})$ can  be evaluated  in an experiment using Eq.(1) as, 
\begin{eqnarray}
W(\theta_{12}) = C
\sum_{i\neq j=0}^{14}  \frac{N_{ij}(\theta_{12})} {\epsilon_{i}(E_1)  \epsilon_{j}(E_2) }, 
\end{eqnarray}
where $C$ is a constant  and the sum is taken over all possible combinations of $(i,j)$ pairs having the angle $\theta_{12}$.  
 In the analysis, for every pair $(i,j)$ of observed $\gamma$ rays, we calculate  $z=cos\theta_{12}$ and  fill the histogram at a position $z$ with a weight $\frac{N_{ij}(\theta_{12})} {\epsilon_{i}(E_1)  \epsilon_{j}(E_2) }$ given by the right-hand side of Eq.(2).  An overall constant $C$ in Eq.(2)  is arbitrary in the present analysis, but if we evaluate the sum on the right hand side of Eq.(2), it should be roughly equal to the number of two  $\gamma$-ray events produced in the target.  
Any deviation from a uniform and  constant distribution of $W(\theta_{12})$ with respect to $cos \theta_{12}$  suggests the existence of an angular correlation between the two  $\gamma$ rays. 

The calculation of the angular correlation function for the two  $\gamma$ rays from  cascade transitions is based  on the electromagnetic theory and quantum numbers conservation as given in Ref.~\cite{Frauenfelder, Biedenharn, Rose}. 
The angular correlation function $W(\theta)$ is conveniently written in terms of  Legendre polynomials as, 
\begin{eqnarray}
W(\theta) = \sum_{\ell =0}^{\ell_{max}}A_{\ell}P_{\ell}(cos\theta), 
\end{eqnarray}
where $P_{\ell}(z)$ is a Legendre polynomial of a degree $\ell$ and $A_{\ell}$ is the coefficient.  
When the detectors for the two $\gamma$  rays are placed (roughly) at cylindrically symmetrical positions from a given source point, this form is simplified  to contain only leading order terms as $\ell$=0, 2 and 4, limiting transitions to dipole and quadrupole types, as
\begin{eqnarray}
W(\theta) = C[1+A_2P_2(cos\theta)+A_4P_4(cos\theta)],  
\label{eq:CorrFun}
\end{eqnarray}
where $C$ is an overall constant. In any experiment, each $\gamma$-ray detector has a finite size and the 
angular correlation function is subject to the correction for the finite size effect or the angular resolution effect~\cite{Rose1, Camp}. If this effect is taken into  account, the coefficients in Eq.(4) are written as, 
\begin{eqnarray}
A_2=A^\prime _2 Q_2\ \ {\rm and}\ \ A_4=A^\prime _4 Q_4, 
\label{eq:FiniteCorr}
\end{eqnarray}
where $Q_2$ and $Q_4$ are the correction factors, and  $A^\prime _2$ and $A^\prime _4$ are the 
coefficients when each detector has a perfect angular resolution, namely $Q_2$=1.0 and $Q_4$=1.0. For the finite angular resolutions, $Q_2$ and $Q_4$ are less than 1.0 and the measured values for $A_2$ and $A_4$ become smaller than the theoretical values for $A^\prime _2$ and $A^\prime _4$. The formula and tabulated values for the coefficients, $A^\prime _2$ and $A^\prime _4$, are given in Ref.~\cite{Biedenharn}. The formula for the correction factors $Q_2$ and $Q_4$ are also given in Ref.~\cite{Rose1, Camp}. \\
\indent
In our experiment, the angular correlation function $W(\theta)$ is analysed using Eqs.(4) and (5) to determine the coefficients $A^\prime _2$ and $A^\prime _4$. Then, the measured values, $A^\prime _2$ and $A^\prime _4$, can be compared with the theoretical values, $A^\prime _2$ and $A^\prime _4$ ~\cite{Biedenharn}. In our ANNRI geometry, the correction factors are calculated to be $Q_2$=0.93$\pm$0.01 ($Q_2$=0.94$\pm$0.01) and $Q_4$=0.77$\pm$0.01 ($Q_4$=0.80$\pm$0.01) for $|cos\theta|>$0.6 ($|cos\theta|<$0.4), respectively. The uncertainty  in the correction factors comes from the uncertainty in the dead layer thickness (1mm) of the Ge crystal~\cite{Utsunomiya, Terada}.

\subsection{Angular correlation of the two $\gamma$ rays from the cascade transition in $^{60}$Co $\beta ^-$ decay}

\begin{figure}[!htbp]
  \begin{center}
	\includegraphics[width=0.8\linewidth]{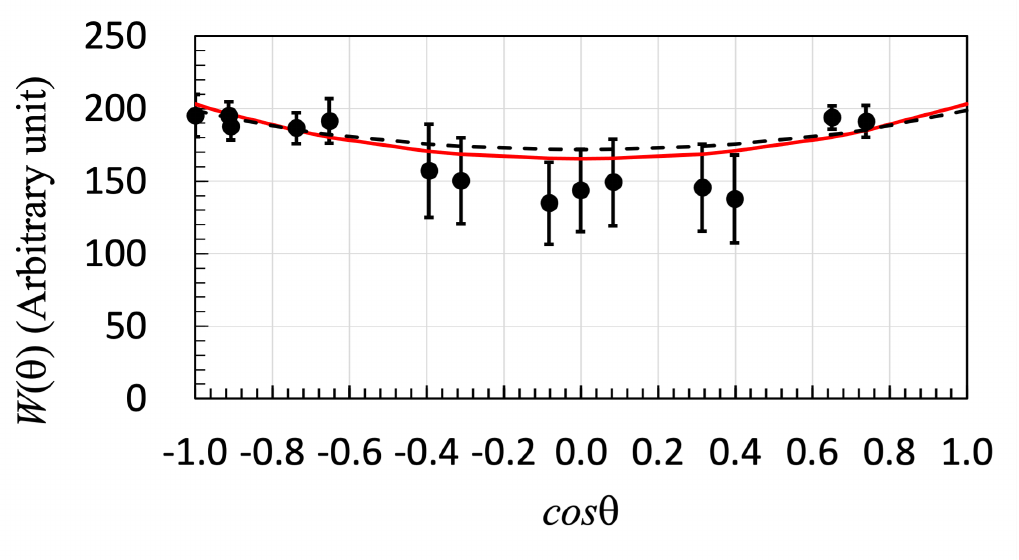}
  \caption{\label{figCo} Measurement of the angular correlation of the two $\gamma$ rays of 1173 keV and 1332 keV from the cascade transition  (2505 keV, $4^+ \rightarrow$ 1332 keV, $2^+ \rightarrow$ 0  keV, $0^+$)  of  $^{60}$Ni in $^{60}$Co $\beta ^-$ decay, The black dashed curve is calculated with the theoretical values for $A^\prime_2$=0.102 and $A^\prime_4$=0.0091 in Eqs.(4) and (5). The red curve is the prediction with $A^\prime_2$=0.15 and $A^\prime_4$=0.0091.  An overall normalization is arbitrary.}
	\end{center}
\end{figure}

Fig.\ref{figCo} shows the angular correlation of the two $\gamma$ rays of 1173 keV and 1332 keV from the cascade transition  (2505 keV, $4^+ \rightarrow$ 1332 keV, $2^+  \rightarrow$  0 keV, $0^+$)  of  $^{60}$Ni from $^{60}$Co $\beta ^-$ decay. We used only the data set of the M2H2 sample. In this analysis, the $^{60}$Co source was set in the target position of the ANNRI detector. 
The predicted values for the coefficients are $A^\prime _2$=0.1020  and $A^\prime _4$=0.0091, respectively. The expected angular correlation is shown in the dashed black curve in Fig.\ref{figCo} 
and it agrees well with data, with $\chi ^2 /dof$=10.5/13. 
If we fit the data using Eqs.(4) and (5) with $A^\prime _2$ being a free parameter and with the fixed value of $A^\prime _4$=0.0091, we obtain $A^\prime _2$=0.15$\pm$0.06 with $\chi ^2 /dof$=9.3/12, which is consistent with the expected value 0.091 within the given uncertainty.  The predicted curve is shown as a red solid curve in Fig.\ref{figCo}. 

\section{Analysis and result}
\subsection{Angular correlation of the two $\gamma$ rays for prominent discrete cascade transitions}

We now study the angular correlation of the two $\gamma$ rays resulting from the prominent discrete cascade transitions of $^{158}$Gd and $^{156}$Gd nuclei. 

\begin{figure}[!t]
  \begin{center}
	\includegraphics[width=0.6\linewidth]{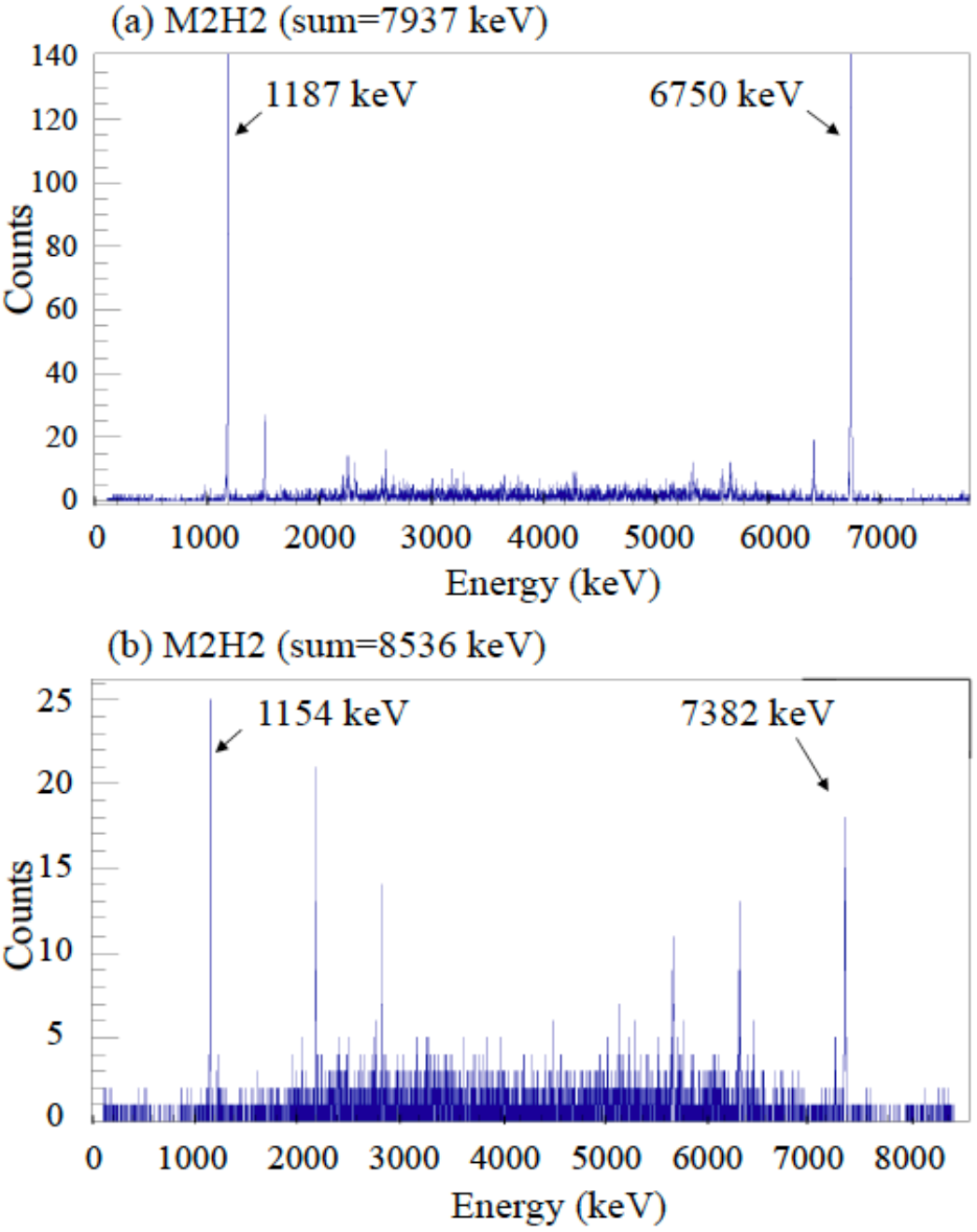}
  \caption{\label{figM2H2} (a) Selection of the two $\gamma$ rays of 6750 keV and 1187 keV in the two-step cascade transition (7937  keV, $2^- \rightarrow$ 1187 keV, $2^+ \rightarrow$ 0 keV, $0^+$)  of $^{158}$Gd. (b) Selection of the two $\gamma$ rays of 7382 keV and 1154 keV in the two-step cascade transition (8536  keV, $2^- \rightarrow$ 1154 keV, $2^+ \rightarrow$ 0 keV, $0^+$)  of $^{156}$Gd.}
	\end{center}
\end{figure}

The process of producing two  $\gamma$ rays  of 6750 keV and 1187 keV, or,  6750 keV and 1107 keV in the cascade transition  of  $^{158}$Gd was already shown in Fig.\ref{fig3}. Figs.\ref{figM2H2}(a) and \ref{figM2H2}(b) exemplify the selection of the two $\gamma$ rays in the M2H2 sample. We show in Fig.\ref{figM2H2}(a) the energy  of the two $\gamma$-rays ($E_1$ and $E_2$), in which the sum $E_1 + E_2$ is equal to 7937 keV within $\pm$25 keV in the M2H2 sample. We select the strongest two  peaks due to 6750 keV and 1187 keV where we observe  almost no random background. The background rate estimation will be described later. 

 For the angular correlation function for the two  $\gamma$ rays  of 6750 keV and 1187 keV in the  two-step cascade transition (7937 keV, $2^- \rightarrow$ 1187 keV, $2^+ \rightarrow$ 0 keV, $0^+$)  of  $^{158}$Gd, the expected coefficients are  $A^\prime _2$=0.25 and $A^\prime_4$=0.
In this cascade transition, the first transition is $E1$ and the second is $E2$. For the cascade transition including an $E1$ transition, $A^\prime _4$ is expected to be 0.0~\cite{Biedenharn, Rose}.  

\begin{figure}[!htbp]
  \begin{center}
	\includegraphics[width=0.8\linewidth]{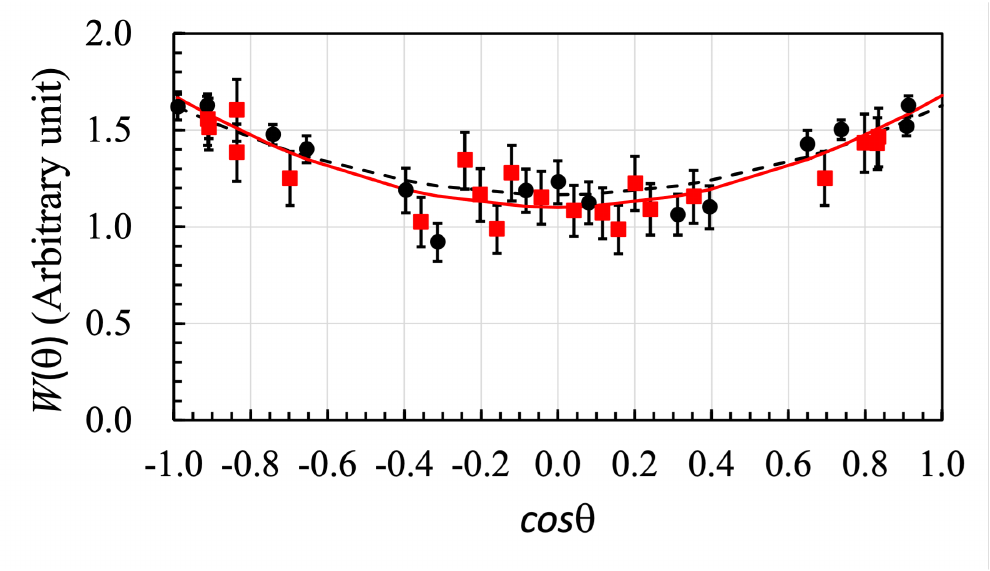}
  \caption{\label{fig6} Measurement of the angular correlation function for the two $\gamma$ rays  of 6750 keV and 1187 keV in the two-step cascade transition (7937  keV, $2^- \rightarrow$ 1187 keV, $2^+ \rightarrow$ 0 keV, $0^+$)  of $^{158}$Gd. The data points  of the  M2H2 sample and the M2H3 sample are plotted in black closed circles and  red closed squares, respectively. The prediction with $A^\prime_2$=0.31 and that with the nominal value $A^\prime_2$=0.25 are shown in the red solid curve and the black dashed curve. Both curves are consistent with the data. }
	\end{center}
\end{figure}

                                        
The angular correlation function measured for these two $\gamma$ rays is shown in Fig.\ref{fig6}. We used two sets of events, namely the  M2H2 sample (black closed circles) and the M2H3 sample (red closed squares).  The data points have been corrected for efficiencies and acceptances according to Eq.(2).  The error bars for all data points are calculated by adding the statistical and systematic uncertainties  in quadrature. The data show a strong angular correlation between the two $\gamma$ rays. 
 If we fit the data using Eqs.(4) and (5) with $A^\prime_2$ being a free parameter and a fixed value of $A^\prime_4$=0.0, we obtain $A^\prime_2$=0.31$\pm$0.03. 
The best fit result with $A^\prime_2$=0.31 and $A^\prime_4$=0.0 is shown in the red solid curve. The agreement between the fit and the data is good ($\chi ^2 /dof  $=31/35). The  best fit values are consistent with the prediction of the expected value 0.25 (shown in black dashed curve) within about 2$\sigma$ level. 

At first glance, the energy distributions shown in  Figs.\ref{fig3} and \ref{fig6}(a) indicate that the background to this sample is negligible. However, we note that there is a chance that the cascade transition of 1107 keV and 80 keV can enter the same crystal, which results in a peak at 1187 keV. Its strength cannot be estimated by the extrapolation of the side-band background rates to the 1187-keV peak. In the following, we  call this probability the coincidence summing probability. We estimated the coincidence summing probability to be about 5$\times 10^{-3}$ by counting the number of events in the 7937-keV peak in  $^{158}$Gd data caused by the coincidence sum of the two $\gamma$ rays of 6750 keV and 1187 keV (7937  keV, $2^- \rightarrow$ 1187 keV, $2^+ \rightarrow $ 0 keV, $0^+$).  Similarly, the 8536-keV peak in  $^{156}$Gd data is caused by  the coincidence sum of the two $\gamma$ rays of 7382 keV and 1154 keV (8536  keV, $2^- \rightarrow$ 1154 keV, $2^+ \rightarrow $ 0 keV, $0^+$). The coincidence summing probability of the 8536-keV peak was found to agree with that of the 7937-keV peak in  $^{158}$Gd data  within 20\%.  For both photopeaks, the direct $M2$ transition of the resonance state  
($2^-$) to the ground state ($0^+$) is strongly suppressed, compared to the $E1$ transition from the resonance state ($2^-$)  to the 1187-keV state ($2^+$, $^{158}$Gd) or to the 1154-keV state ($2^+$, $^{156}$Gd)~\footnote{We mistakenly listed the intensity of the  7937-keV peak as 0.55$\pm$0.03($\times10^{-2}$\%) in the Table 1 of our previous publication~\cite{Hagiwara}. We used this intensity of the 7937-keV peak to estimate the coincidence summing probability in this paper.}. 
The coincidence summing probability to the 1187-keV peak is less than 1\%. In addition, we checked all possible pairs of two $\gamma$ rays in the M2H2 sample with a coincidence sum that evaluates to 6750 keV. Such pairs of the two $\gamma$ rays include 5903 keV and 847 keV (7937  keV, $2^- \rightarrow$ 2034 keV, $3^+ \rightarrow $ 1187 keV, $2^+$) and 5784 keV and 966 keV (7937  keV, $2^- \rightarrow$  2153 keV, $2, 3^+ \rightarrow $ 1187 keV, $2^+$). We estimated the coincidence summing probability to be about 1.5\% of the total  number of counts in the single photopeak of 6750 keV. The coincidence summing effect to the angular correlation function is negligible. 

\begin{figure}[!htbp]
  \begin{center}
	\includegraphics[width=10.0cm,scale=1.0]{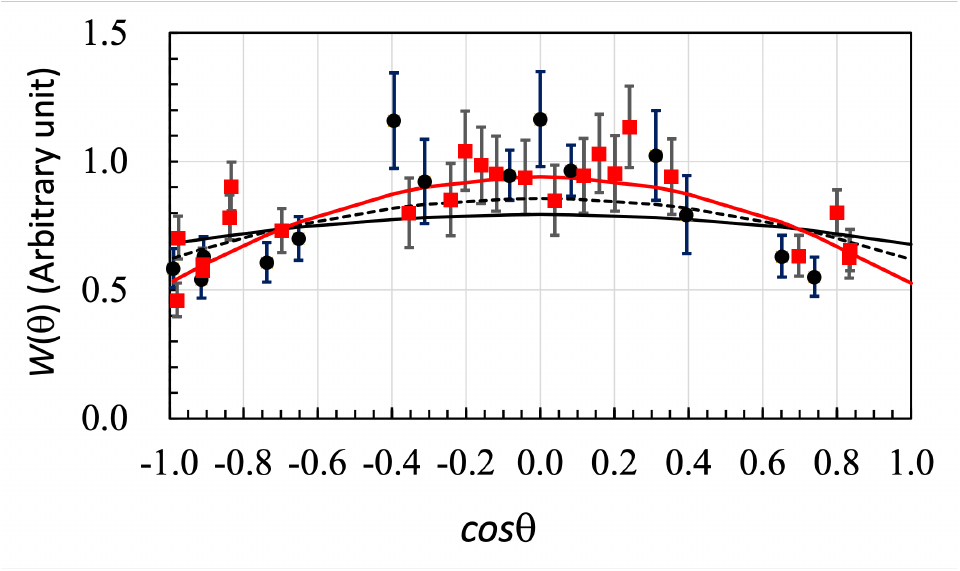}
  \caption{\label{fig7}Measurement of the angular correlation between two $\gamma$ rays of 6750 keV and 1107 keV in the two-step cascade
  ($7937 \  keV, 2^- \rightarrow 1187 \ keV, 2^+ \rightarrow 80 \ keV, 2^+$)  of  $^{158}$Gd. The data points  of the  M2H2 sample and the M2H3 sample are plotted in black closed circles and  red closed squares, respectively.
  The predictions with $A^\prime_2$=-0.11 ($\delta$=-9.0),  $A^\prime_2$=-0.22 ($\delta$=-1.5) and $A^\prime_2$=-0.37  are drawn in black solid curve, black dashed curve and red solid curve, respectively.}
	\end{center}
\end{figure}

\begin{figure}[!htbp]
  \begin{center}
	\includegraphics[width=0.8\linewidth]{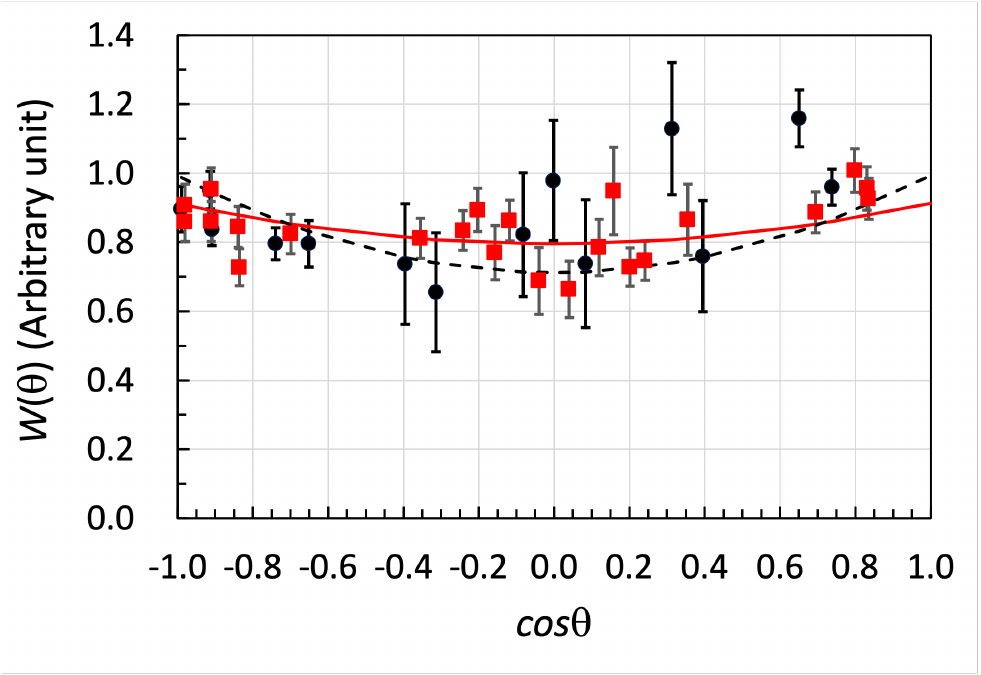}
  \caption{\label{fig8}Measurement of the angular correlation between two $\gamma$ rays of 7382 keV and 1154 keV  in the prominent two-step cascade transition (8536  keV, $2^- \rightarrow$ 1154 keV, $2^+ \rightarrow$ 0 keV, $0^+$) for $^{156}$Gd. The data points  of the  M2H2 sample and the M2H3 sample are plotted in black closed circles and  red closed squares, respectively. The predictions with $A^\prime_2$=0.10 (best fit, red solid curve) and that with the nominal value $A^\prime_2$=0.25 (black dashed curve) are shown, }
	\end{center}
\end{figure}

Next, the result for the angular correlation between the  two $\gamma$ rays of 6750 keV and 1107 keV in the cascade transition from $^{158}$Gd (7937 keV, $2^-$) are shown in Fig.\ref{fig7}. Note that  the 80-keV $\gamma$ ray in Fig.\ref{fig3}(right) cannot be detected by ANNRI since it is below our experimental threshold. 
The theoretical prediction for the angular correlation function is estimated for  $E1$-$E2$ transition and for $E1$-$M1$ transition as, 
\begin{eqnarray}
A^\prime _2&=&-0.054, \ A^\prime _4=0\  {\rm for\ a\ pure}\  E1-E2  {\rm\ transition \ and,} \nonumber \\
A^\prime _2&=&0.175, \ A^\prime _4=0\  {\rm for\ a\ pure}\  E1-M1  {\rm \ transition.}
\end{eqnarray}
Fig.\ref{fig7} clearly shows a negative value for $A^\prime_2$. If we fit the data using Eqs.(4) and (5) with $A^\prime_2$ being a free parameter and with a fixed value of $A_4$=0.0, we obtain $A^\prime_2$=-0.37$\pm$0.04.  The agreement between the fit and the data is good  with $\chi ^2 /dof  $=38/35. This fit value $A^\prime_2$=-0.37$\pm$0.04 is not consistent with the prediction of either 
 a pure $E1$-$E2$ transition, or a  pure $E1$-$M1$ transition. The best fit is shown as the red solid curve in Fig.\ref{fig7}.  
The previous measurement of the transition (1187 keV, $2^+ \rightarrow$ 80 keV, $2^+$) was performed in a Coulomb excitation experiment and it reported a mixture of $E2$ and $M1$ transitions with the mixture parameter $\delta$=-9.0$\pm$1.5~\cite{McGowanGd, NDSGd158}, where $\delta$ is defined as the ratio of $E2$ to $M1$ transition~\cite{Frauenfelder, Biedenharn, Arns}. It is noted that the angular distribution is expected to show an interference effect caused by the mixture of two multipoles in a single transition. The coefficient $A^\prime_2$ of the angular correlation function can be calculated~\cite{Frauenfelder, Biedenharn, Arns} when the transition is mixed with a mixture parameter $\delta$ and it is given as
\begin{eqnarray}
A^\prime _2=\frac{0.175+0.510\delta-0.0536\delta^2}{1+\delta ^2},  
\end{eqnarray}
where $A^\prime _4$=0, since the first transition is $E1$. The value of $A^\prime_2$ is estimated by Eq.(7) to be -0.11$\pm$0.01 for  the previously reported value $\delta$=-9.0$\pm$1.5 and its prediction is drawn in the black  solid curve in Fig.\ref{fig7}. Our data are inconsistent with this value. 

If we fit the data with $\delta$ as a free parameter, we obtain $\delta=-1.5^{+1.5}_{-0.5}$, which gives $A^\prime_2$=-0.22 from Eq.(7). The prediction is barely consistent with the data. We note that the previous measurement of the transition was measured by comparing the ratio of the 1107-keV rate at two different angles (0$^\circ$  and  90$^\circ$) with respect to the beam~\cite{McGowanGd}. Systematic effects in the previous and the present experiment which measured  the angular correlation function at all angles are rather different. 
The background to our angular correlation analysis due to the coincidence summing effect is at the same level as that of Fig.\ref{fig6} and is estimated to be negligible. 

Fig.\ref{fig8} shows the angular  correlation  function for the prominent two $\gamma$ rays of the 7382 keV and 1154 keV  in the two-step cascade transitions (8536 keV, $2^- \rightarrow$ 1154 keV, $2^+ \rightarrow$ 0 keV, $0^+$) for $^{156}$Gd. We show in Fig.\ref{figM2H2}(b) the energy  of the two $\gamma$-rays, in which the sum $E_1 + E_2$ is equal to 8536 keV within $\pm$25 keV in the M2H2 sample. We select the two  peaks due to 7382 keV and 1154 keV unambiguously.
For this case, the theoretical prediction for the angular correlation function 
is the same as for the  two-step cascade transition (7937  keV, $2^- \rightarrow$ 1187 keV, $2^+ \rightarrow$ 0 keV, $0^+$)  of  $^{158}$Gd, but the result shown in Fig.\ref{fig8} looks rather different from that of Fig.\ref{fig6}. 
 If we fit the data using Eqs.(4) and (5) with $A^\prime_2$ being a free parameter and with the fixed value of $A^\prime_4$=0.0, we obtain $A^\prime_2$=0.10$\pm$0.04 and the quality of the fit is relatively poor with $\chi ^2 /dof  $=58/35. The prediction with theoretical value $A^\prime_2$=0.25 is also shown as the black dashed curve. 

 We now consider the background levels to each peak of 1154 keV and 7382 keV. The background for the 1154-keV peak caused by the coincidence sum of 1075 keV and 79 keV is estimated to be 0.5\%. 
 We also checked all possible pairs of two $\gamma$ rays in the M2H2 sample whose coincidence sum results in a peak at 7382 keV. We found that  the number of pairs is  more by about a factor of 5 than that for 6750 keV. The pairs of the two $\gamma$ rays are  6345 keV and 1037 keV, which are produced in a cascade transition (8536 keV, $2^- \rightarrow$ 2191 keV, $2^+ \rightarrow$ 1154 keV, $2^+$),  6745 keV and  637 keV, 6427 keV and 955  keV, 6381 keV and 901 keV,  and  6319 keV and  1063  keV. We estimate the coincidence summing probability of all pairs to be about 7.7\% of the total  number of a single photopeak of 7382 keV. Thus, the background to the pairs of the two $\gamma$ rays of 1154 keV and 7382 keV is estimated to be 8.2$\pm$2.0\%.
 Those backgrounds may have smeared the  angular correlation function in addition to the poorer statistics of this sample than that of  the $^{158}$Gd data, as seen in Fig.\ref{figM2H2}.



\subsection{Angular correlation of the two $\gamma$ rays for the continuum}

We also studied the angular correlation of the two $\gamma$ rays emitted in the continuum transitions. In this analysis, we used  only the two $\gamma$ rays from the M2H2 sample for simplicity. 

\begin{figure}[!htbp]
  \begin{center}
	\includegraphics[width=15.0cm,scale=1.0]{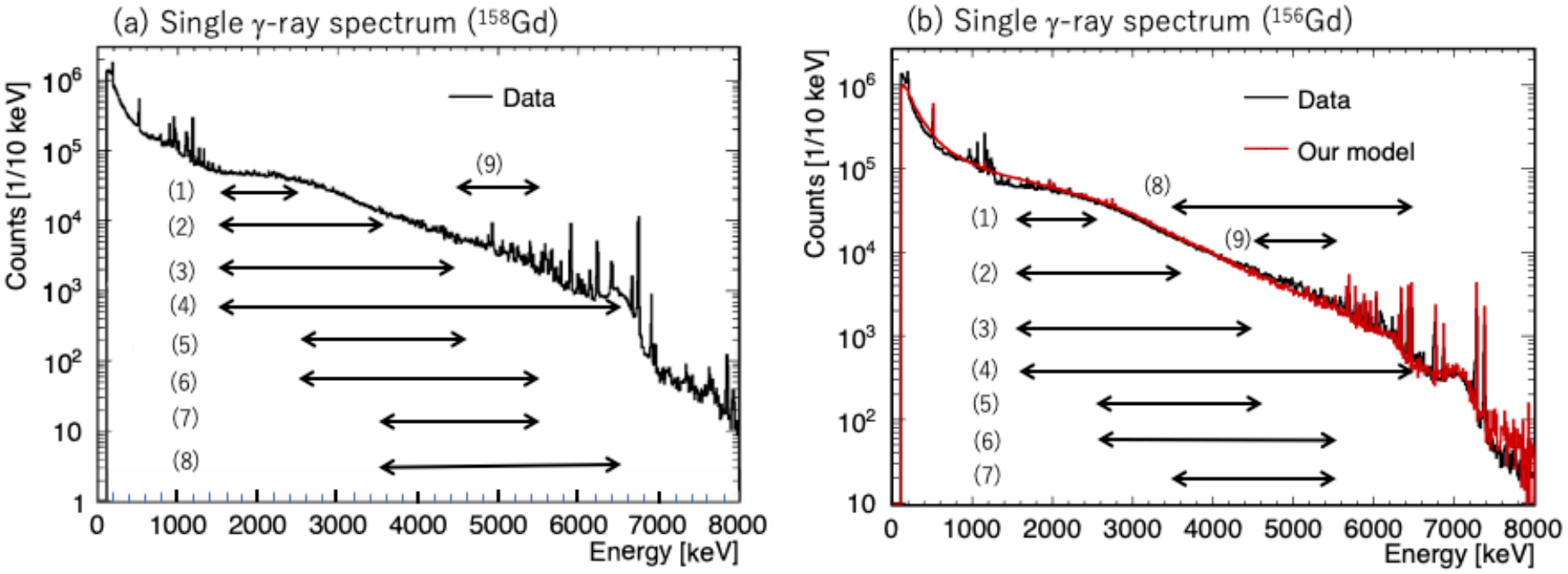}
  \caption{\label{fig9}Energy ranges (1)-(9)($E_1$-$E_2$) of the two $\gamma$ rays are shown in
  arrows over the single energy spectrum of (a) $^{158}$Gd and (b) $^{156}$Gd. Fig.\ref{fig9}(a) and Fig.\ref{fig9}(b) were taken from Fig.4 (Ref.~\cite{Hagiwara}) and Fig.9(left) (Ref.~\cite{Tanaka}) of our previous publications, respectively. }
	\end{center}
\end{figure}

In addition, we required that the energies of the two $\gamma$ rays are within nine predefined energy ranges that avoid the energies of the strong discrete photopeaks that we have investigated above. 
The nine energy ranges ($a<E_1, E_2<b)$ are chosen as follows: $a-b$ MeV=(1) 1.5-3.5 MeV, (2) 1.5-3.5 MeV,  (3) 1.5-4.5 MeV, (4) 1.5-6.5 MeV, (5) 2.5-4.5 MeV, (6) 2.5-5.5 MeV, (7) 3.5-5.5 MeV, (8) 3.5-6.5 MeV, and (9) 4.5-5.5 MeV. They have been superimposed on the $\gamma$-ray spectra of $^{157}$Gd($n, \gamma$)  and $^{155}$Gd($n, \gamma$) reactions in Fig.\ref{fig9}. Fig.\ref{fig9}(a) and Fig.\ref{fig9}(b) were taken from Fig.4 (Ref.~\cite{Hagiwara}) and Fig.\ref{fig12}(left) (Ref.~\cite{Tanaka}) of our previous publications, respectively.  

\begin{figure}[!htbp]
  \begin{center}
 \includegraphics[width=4.5cm,scale=1.0]{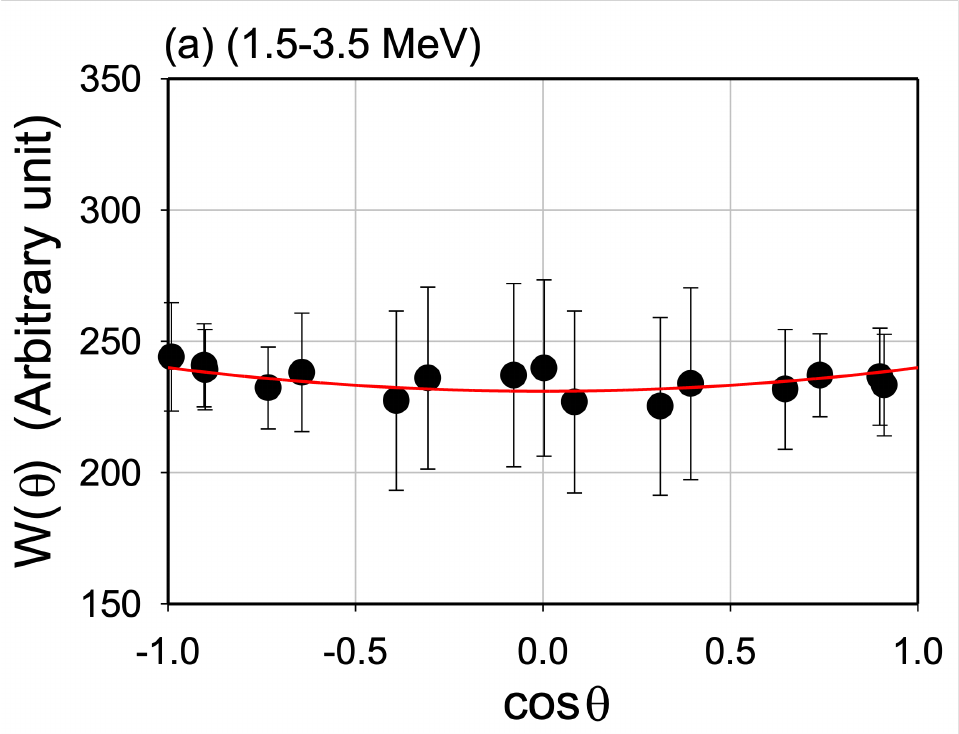}
 \includegraphics[width=4.5cm,scale=1.0]{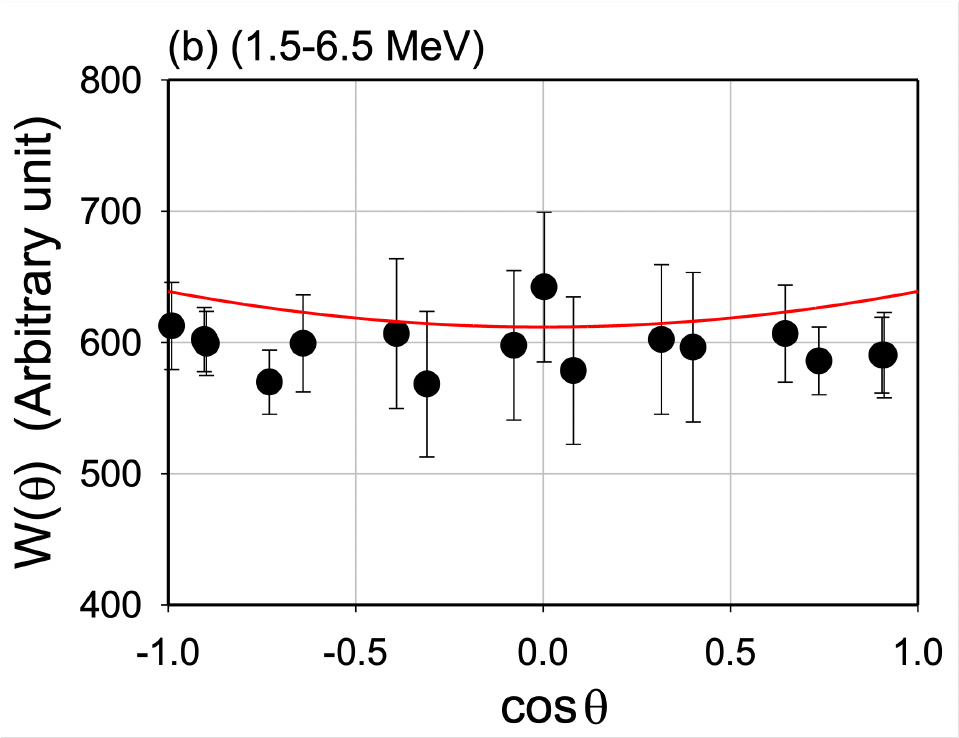}
 \includegraphics[width=4.5cm,scale=1.0]{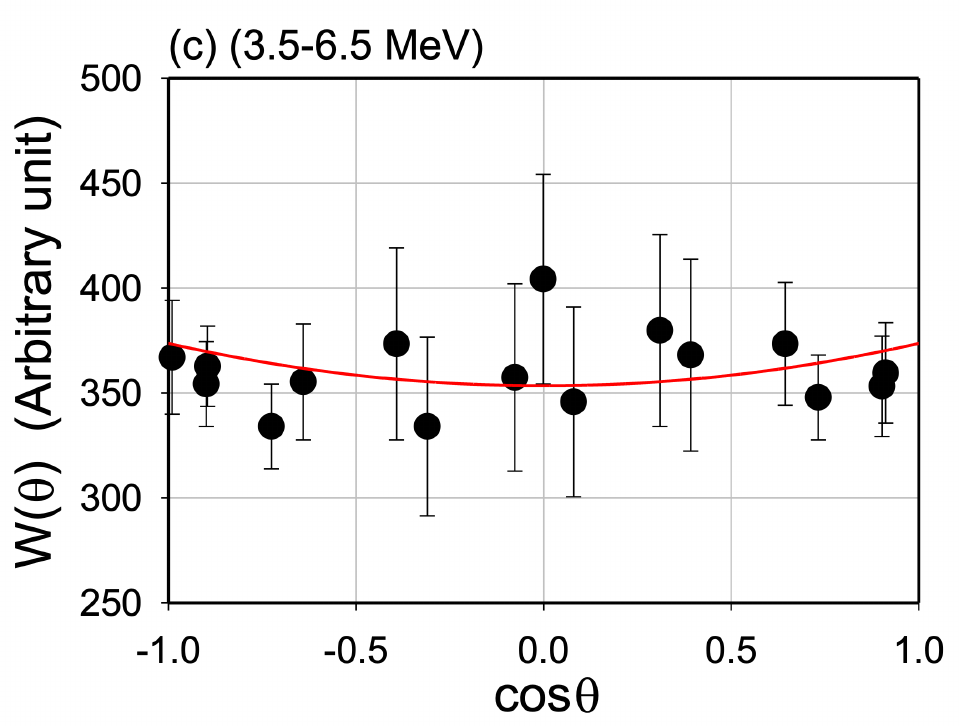}
  \caption{\label{fig10}Measurement of the angular correlation between two $\gamma$ rays ($E_1$ and $E_2$) from the continuum for $^{158}$Gd decays. The energy range for $E_1$ and $E_2$ is (a)(1.5 MeV, 3.5 MeV), (b) (1.5 MeV, 6.5 MeV), and (3.5 MeV, 6.5 MeV).  The red lines show the best fit, which is consistent with a flat distribution.}
	\end{center}
\end{figure}

\begin{figure}[h]
  \begin{center}
 \includegraphics[width=4.5cm,scale=1.0]{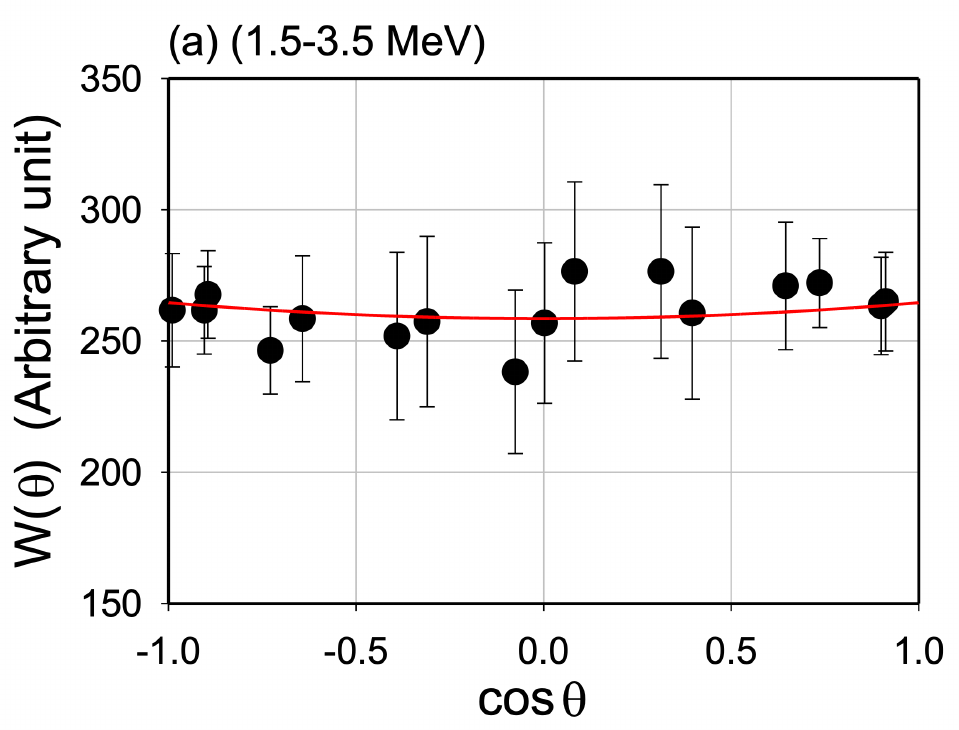}
 \includegraphics[width=4.5cm,scale=1.0]{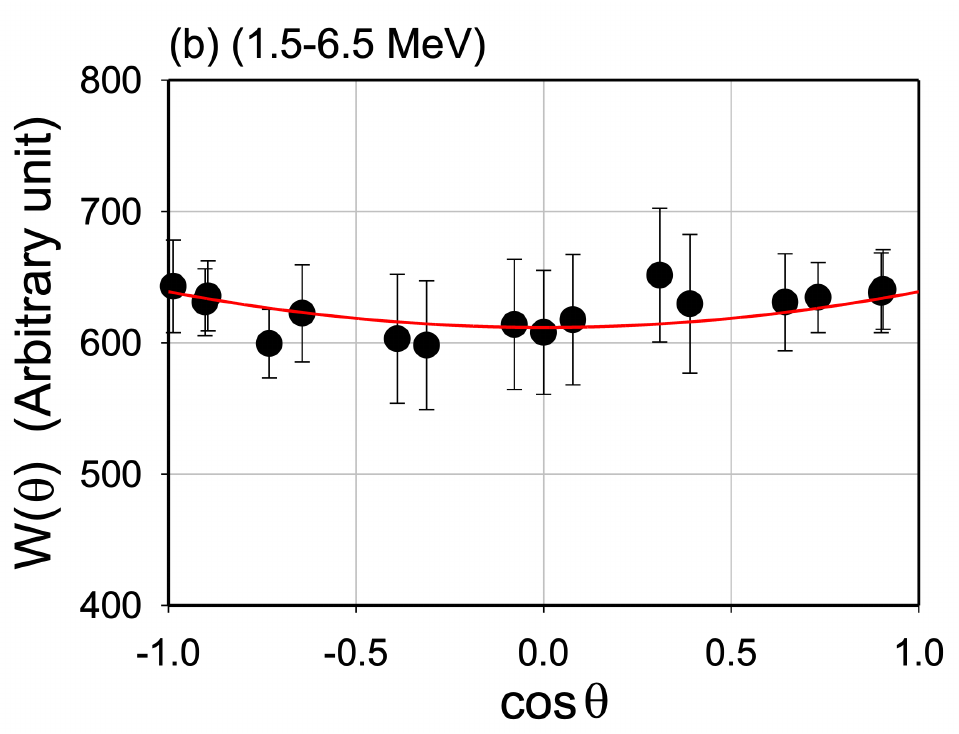}
 \includegraphics[width=4.5cm,scale=1.0]{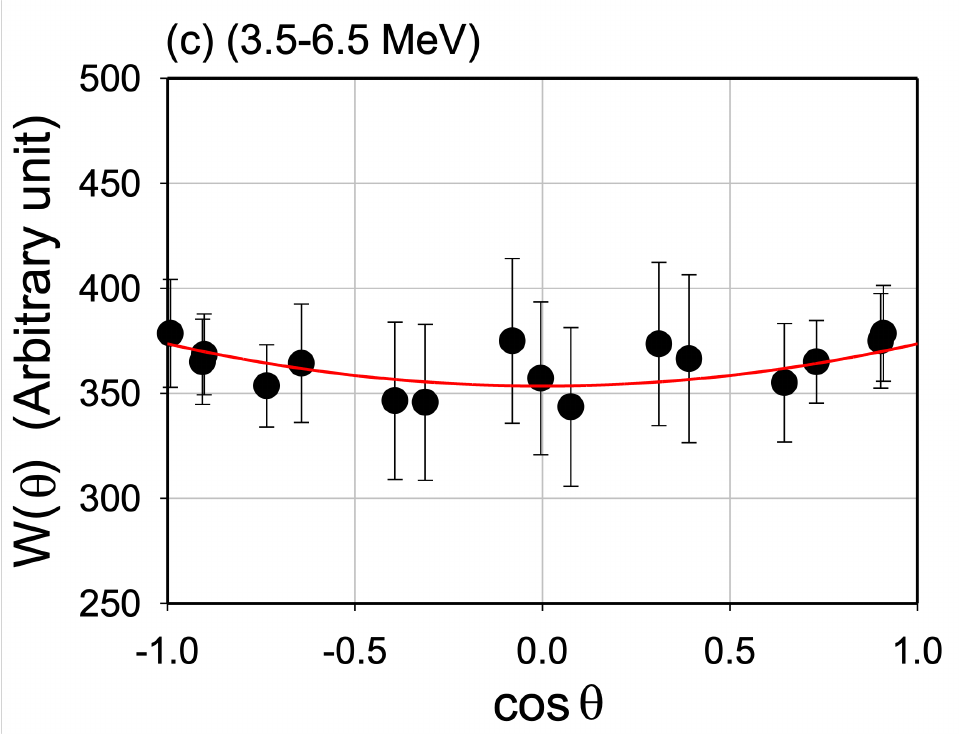}
  \caption{\label{fig11}Measurement of the angular correlation between two $\gamma$ rays ($E_1$ and $E_2$) from the continuum for $^{156}$Gd decays. The energy range for $E_1$ and $E_2$ is (a)(1.5 MeV, 3.5 MeV), (b) (1.5 MeV, 6.5 MeV), and (3.5 MeV, 6.5 MeV).  The red lines show the best fit, which is consistent with a flat distribution.}
	\end{center}
\end{figure}

\begin{figure}[h]
  \begin{center}
    \includegraphics[width=7.0cm,scale=1.0]{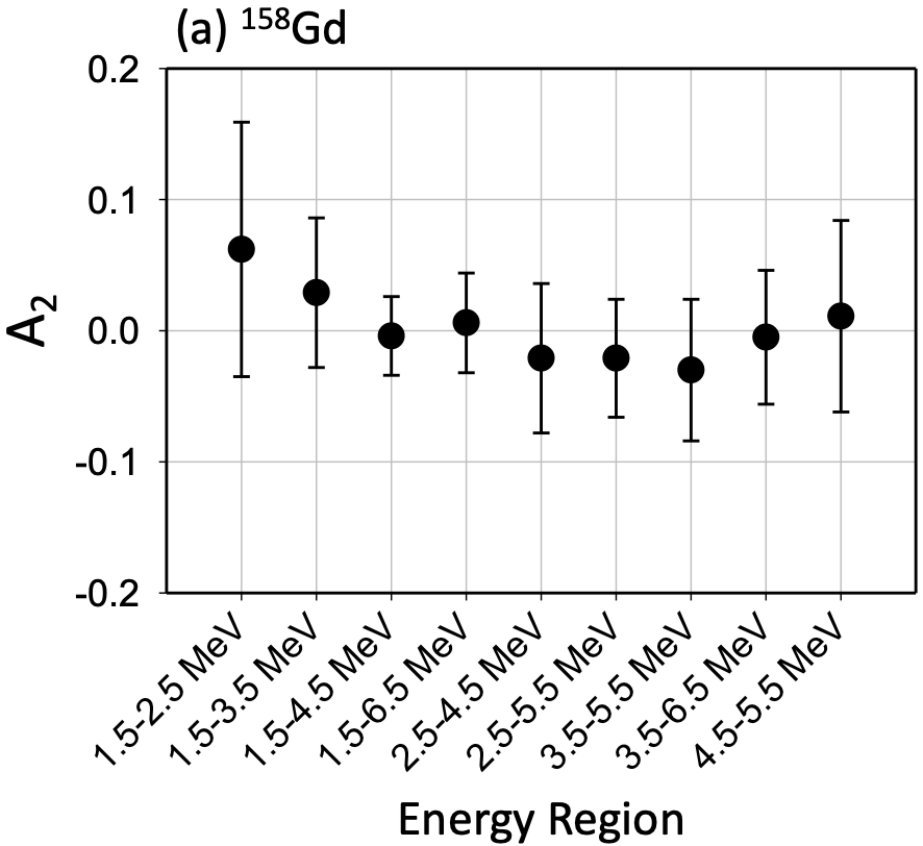}
    \includegraphics[width=7.0cm,scale=1.0]{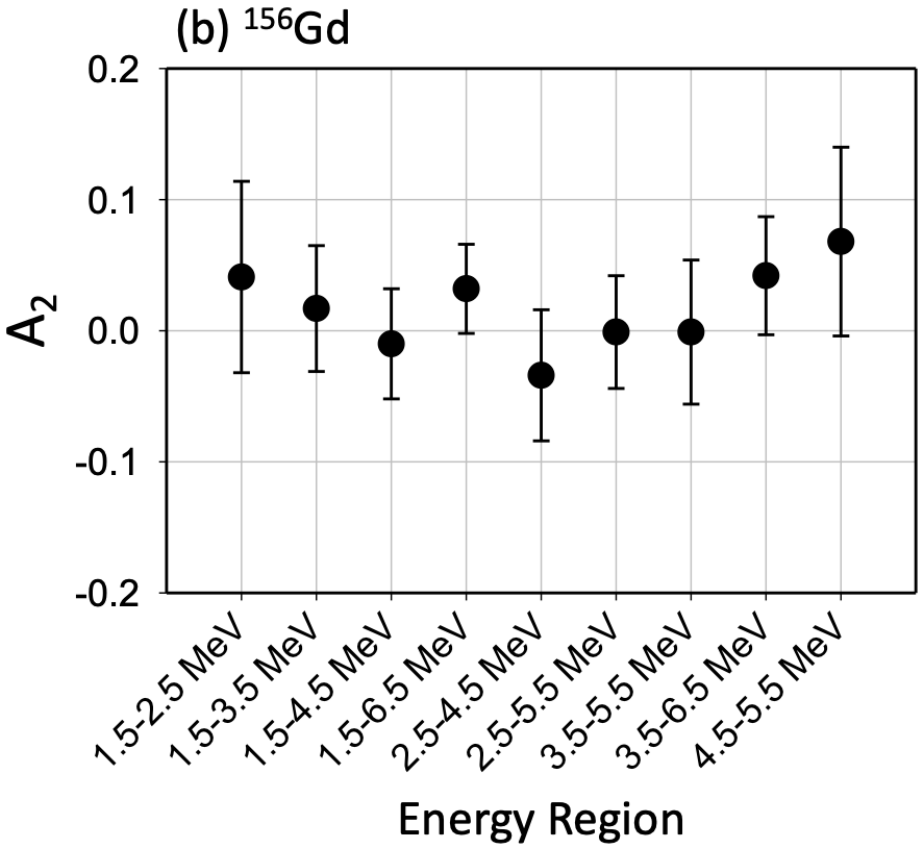}
    \caption{Coefficient $A_2$ ((a) $^{158}$Gd and (b) $^{158}$Gd) of the angular correlation function plotted for various energy ranges ($E_1$-$E_2$) of the two $\gamma$ rays in the continuum. The angular corrrelation function is assumed to be in the form of $W(\theta) =C[1+A_2P_2(cos\theta)]$. $A_2$=0 means no correlation.}
    \label{fig12}
  \end{center}
\end{figure}

Figs.\ref{fig10} and \ref{fig11} present exemplarily the results of angular correlation functions for the energy ranges (2), (4) and (7) for the  $^{158}$Gd and $^{156}$Gd data. The error bars displayed in the figures include  both  statistical and systematic uncertainties. 
We analysed the angular correlation functions for all energy ranges, assuming a form $W(\theta) =C[1+A_2P_2(cos\theta)]$, where a constant $C$ and  the coefficient $A_2$ are  the free  parameters. The results for the  coefficient $A_2$ for all the energy ranges are shown in Fig.\ref{fig12}, where the uncertainty of the coefficient $A_2$ is determined by $\chi ^2$= $\chi ^2_{minimum}+1.0$.    
The values of the coefficient $A_2$ in any energy range are consistent with 0 within a few \%. Hence, we observe no significant angular correlations for any combinations of two  $\gamma$ rays from the continuum. 


\section{Summary and discussion}

Using the ANNRI Ge spectrometer setup at J-PARC, we have studied  for the first time the angular correlations between the two $\gamma$ rays emitted from $^{155}$Gd and $^{157}$Gd targets after capture of thermal neutrons. 

We have shown that the angular correlation functions between the two prominent $\gamma$ rays  produced in the strong two-step cascade transitions from the resonance state can be described with the functional form of Eqs.(4) and (5) predicted by electromagnetic theory~\cite{Frauenfelder, Biedenharn, Rose}. 
For the angular correlation function for the two  $\gamma$ rays  of 6750 keV and 1187 keV in the  two-step cascade transition (7937 keV, $2^- \rightarrow$ 1187 keV, $2^+ \rightarrow$ 0 keV, $0^+$)  of  $^{158}$Gd, our data shown in Fig.\ref{fig6} are consistent with the prediction within 2$\sigma$ level. The background to the angular correlation function is  negligible. 

Next,  we showed in Fig.\ref{fig7}  the angular correlation function for the two $\gamma$ rays of the 6750 keV and 1107 keV in the cascade  (7937  keV, $2^- \rightarrow$ 1187 keV, $2^+ \rightarrow$ 80 keV, $2^+$) and  compared it with the prediction of the electromagnetic theory.  The best fit value to our data is not consistent with the prediction of either a pure $E1-E2$ transition, or a  pure $E1-E1$ transition. The previous measurement of the transition (1187 keV, $2^+ \rightarrow$ 80 keV, $2^+$) reported a mixture of $E2$ and $M1$ transitions with a mixing parameter $\delta$=-9.0$\pm$1.5~\cite{McGowanGd}. Instead, our fit to the angular correlation function  results   $\delta=-1.5^{+1.5}_{-0.5}$, corresponding to $A^\prime_2$=-0.22 from Eq.(7). Our result is not consistent with the previous  measurement.  
We note that the previous measurement and the present experiment which measured  the angular correlation function at all angles use   different experimental methods. Further measurements will be necessary.  
The background to the angular correlation analysis in Fig.\ref{fig7} due to coincidence summing effect is again estimated to be negligible. 

We also studied the angular  correlation  function for the prominent two $\gamma$ rays of the 7382 keV and 1154 keV  in the  two-step cascade transitions (8536 keV, $2^- \rightarrow$ 1154 keV, $2^+ \rightarrow$ 0 keV, $0^+$) for $^{156}$Gd in Fig.\ref{fig8}. For this case, the theoretical angular correlation function 
should be the same as the  two-step cascade transition (7937  keV, $2^- \rightarrow$ 1187 keV, $2^+ \rightarrow$ 0 keV, $0^+$)  of  $^{158}$Gd, but the result shown in Fig.\ref{fig8} are different from Fig.\ref{fig6}. 
 We also checked all possible pairs of two $\gamma$ rays in the M2H2 sample whose coincidence sum results in a peak at 7382 keV and found that  the number of pairs is  more by a factor of 5 than for 6750 keV. We estimated the coincidence summing probability of all pairs to be about 7.7\% of the total  number of a single photopeak of 7382 keV. Thus, the background to the pairs of the two $\gamma$ rays of 1154 keV and 7382 keV is estimated to be 8.2$\pm$2.0\%. 
 Those background may have smeared the  angular correlation function in addition to the poorer statistics of this sample than that of  the $^{158}$Gd data (Fig.\ref{figM2H2}). 

Next, we have studied the angular correlations between two  $\gamma$ rays produced from the continuum transitions, assuming that the angular correlation can be written in a form  $W(\theta) =C[1+A_2P_2(cos\theta)]$. We found that the value of the coefficient $A_2$ is consistent with 0 within uncertainties, typically 0.05 and less than 0.1, as shown in Fig.\ref{fig12}. Hence, we found no angular correlations, for any two  $\gamma$ rays in the continuum for energies below 6.5 MeV. 

This result agrees with our expectations since we picked  random pairs of two $\gamma$ rays in the cascade transition and excluded the prominent strong photopeaks from the pairs. 
The mean multiplicity of  $\gamma$ rays produced in the neutron capture reaction is about 5  for $E_{\gamma}>$0.2 MeV. Since we pick a random pair of two $\gamma$ rays in the continuum, the probability that the same pair is selected from the definite spin-parity states must be very small and, as a  result, we expect that they show no angular correlations. 
We note that the $\gamma$ rays from the continuum represent approximately 93{\%} (97\%) of $\gamma$ rays produced in the thermal neutron capture of  $^{\rm 157}$Gd(n, $\gamma$)  reaction  ($^{\rm 155}$Gd(n, $\gamma$) reaction) for $E_{\gamma}>$0.11 MeV~\cite{Hagiwara, Tanaka}.

In summary, our study of the angular correlation for both  the  two $\gamma$ rays from the strong two-step cascade transition and for the randomly chosen  two $\gamma$ rays in continuum is an important information for the ongoing and future experiments using gadolinium for neutrons tagging in a liquid-scintilator detector or in a water-Cherenkov detector. 



\section*{Acknowledgement}
\label{sec:Acknowledgements}

This work is supported by the JSPS Grant-in-Aid for Scientific Research on Innovative Areas 
(Research in a proposed research area) No. 26104006 and also by the JSPS Grant-in-Aid for Scientific Research (C) No. 20K03989. It benefited from the use of the 
neutron beam of the JSNS and the ANNRI detector at the Materials and Life Science Experimental 
Facility of the Japan Proton Accelerator Research Complex.

\end{document}